\input{aipcheck.tex}

\documentclass{aipproc}

\layoutstyle{6x9}

\SetInternalRegister\hbadness{8000} 

\newcommand\doingARLO[2][]{%
  \ifx\mmref\undefined #1\else #2\fi
}

\begin{document}

\title [X-ray flares]
      {X-ray flares and their relation to the inner engine activity}


\author{Davide Lazzati, Brian C. Morsony, Rosalba 
Perna, and Mitchell C. Begelman}
{address={JILA - University of Colorado, 440 UCB, Boulder, CO
80309-0440, USA}, email={lazzati@colorado.edu}, }


\copyrightyear  {2008}

\begin{abstract}
Flares overlaid on the smooth power-law decay of Swift X-ray
afterglows are rather common, appearing in roughly half the observed
light curves. They are a manifestation of the late time activity of the
inner engine, since their temporal evolution is too fast to be linked
to activity taking place in the external shock blastwave. In this
paper we show that the energy emitted in the form of flares decreases
with time as a power-law. We discuss several possibilities in which
the flares can be powered and the source of the observed
variability. We show that late time accretion from a disk can provide
the necessary energy input in both classes of short duration and long
duration gamma-ray bursts.
\end{abstract}

\date{\today}

\maketitle

\section{Introduction}

Approximately half of the X-ray afterglows observed by Swift display
one or more X-ray flares between few hundreds to several tens of
thousand of seconds after the end of the prompt emission
\cite{chinca07,falcone07}. Albeit with some exception, flares are
observed only in the X-ray telescope data, and are therefore much
harder than the afterglow emission.

An short-time increase in the X-ray afterglow flux can have various
causes. These can be broadly distinguished in two classes. In the
first class we put events taking place in the external shock
blastwave. In the second class we put events due to late activity of
the inner engine, i.e., radiation produced by electrons trailing the
afterglow blastwave, released at a later stage by the inner
engine. These two possibilities can be distinguished through the time
behavior of the flares. In \S~2 we will show that when a flare is
observed at a time $T$ after the burst trigger, it means that the
inner engine had an episode of emission at a time very close to
$T$. Such conclusion implies that the flares can tell us precious
information on the functioning of the inner hidden engine at very long
times, when its luminosity has decreased several orders of
magnitude. To start addressing this issue, in \S~3 we compute the
amount of energy that is released by the inner engine in the form of
flares as a function of the time elapsed since the prompt emission and
we elaborate on possible sources for the late energy injection. In
\S~4 we discuss the mechanism by which the energy released from the
engine is observed in episodic flares and not as a continuous
output. Finally, in \S~5 we summarize and discuss our results.

\section{Internal or external?}

\begin{figure}
\includegraphics[width=0.5\textwidth]{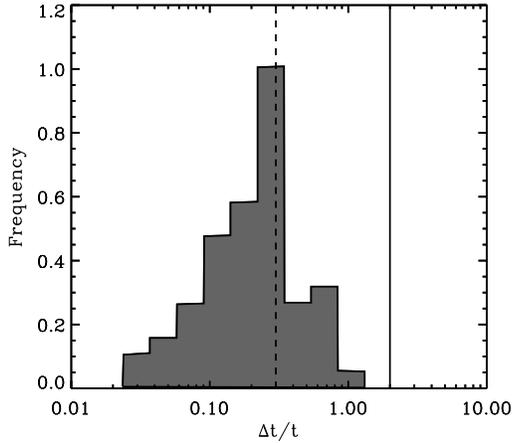}
\caption{{Histogram of the duration of flares over 
the time of the peak (data from \cite{chinca07}). The vertical solid
line shows the minimum duration of a flare from external shock
activity. The dashed line shows the limit for inner engine activity
immediately after the end of the prompt phase.}
\label{fig:flares}}
\end{figure}

Consider a spherical blastwave, for simplicity as a sphere
concentrated on the external shock. Even if the whole sphere suddenly
brightens and dims again in a very short time, the observer at
infinity would see a flare with a long duration. This is due to the
curvature of the emitting surface and to the finite speed of
light. Due to special relativistic effects, the duration of the flare
depends on the spectral energy distribution and on the dynamics of the
blastwave expansion. Lazzati \& Perna \cite{lazzperna07} computed the
minimum duration of a flare as
\begin{equation}
\frac{\Delta{t}}{t} \ge \left(2^{\frac{1}{\eta+2}}-1\right)(4\alpha+2)
\label{eq:limit}
\end{equation}
where $\Delta{t}$ is the FWHM of the pulse, $t$ the peak time, $\eta$
is the spectral index (under the assumption the spectrum is well
described by a power-law), and $\alpha$ a parameter specified by the
dynamics of the blastwave (either radiative or adiabatic) and the
properties of the interstellar medium. For an adiabatic blastwave
expanding into a uniform medium and with a power-law spectral index of
$-1$ ($F(\nu)\propto\nu^{-1}$), we obtain
$\Delta{t}/t>2$. Figure~\ref{fig:flares} shows the distribution of the
observed $\Delta{t}/t$ \cite{chinca07} compared to the limit of
Eq.~\ref{eq:limit}. Not a single flare has a duration that satisfies
the above condition.

The flares must therefore be related to engine activity. A very
important question remains. Are the flares observed at a late time
because the engine is active at late time or are they due to slower
material ejected at the end of the prompt phase with a small Lorentz
factor?  Lazzati \& Perna \cite{lazzperna07} show that also emission
produced by material ejected immediately after the prompt phase has to
satisfy a constraint on the duration that reads $\Delta{t}/t\ge 0.3$
(for a spectral index of $-1$. This constraint is shown in
Fig.~\ref{fig:flares} as a dashed line. Even though some 30 per cent
of the flares can be explained in this way, 70 per cent of them do
require that the inner engine ejected energy at a time immediately
before the flare itself is seen.

\section{Energy in flares}

\begin{figure}
\includegraphics[width=0.5\textwidth]{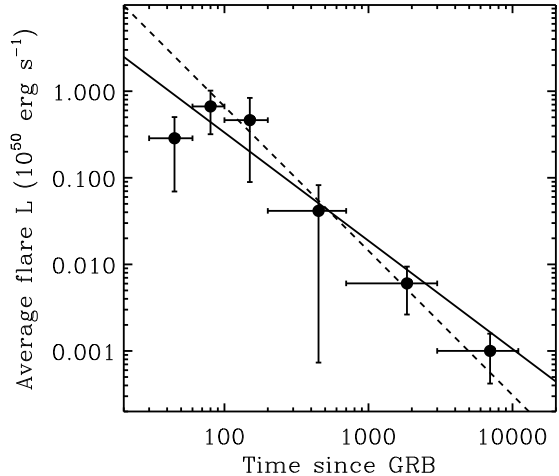}
\caption{{Average light curve in flares (integrated over a 
time much longer than the flare duration) as a function of the
comoving time since the GRB trigger.  The solid line and the dashed
line show the energy supply for a draining thin disk and a fallback
supernova, respectively. Data from \cite{lazz08}.}
\label{fig:curve}}
\end{figure}

In the previous paragraph we showed that the time at which the flare
is observed coincides, in most cases, with the time at which the
energy was ejected from the central engine. This conclusion allows us
to construct an average light curve of flares that tells us how much
energy is released from the central engine as a function of time
\cite{lazz08}. To this aim, we selected a sample of 10 GRBs
from \cite{falcone07} for which the redshift is known. They all have
at least one flare, for a total of 24 flares. The average flare light
curve is shown in Fig.~\ref{fig:curve}. The result is obtained by
assuming that the opening angle of the flares is constant or, at
least, that there is no correlation between the jet opening angle and
the time at which the engine releases the energy to produce a flare,
nor a consistent evolution of the opening angle within a single GRB.

On average, the engine of GRBs releases a tail of energy extending to
several hours (in comoving time) as a power-law of index $\sim-1.5$
\cite{lazz08}. What is the energy supply that is used to power
such tail? In a black hole (BH) accretion disk system there are, at
least, two possibilities. First, fall-back material from the supernova
explosion can rain onto the draining disk, providing an extra supply
of matter to accrete. Chevalier \cite{cheva} showed that such
fall-back has a power-law behavior with index $-5/3$. The dashed line
in Fig.~\ref{fig:curve} shows a $-5/3$ decay overlaid on the data. It
provides a good fit. Alternatively, the disk itself can supply energy
for a long time, as a consequence of the spreading of the disk to
large radii in order to conserve angular momentum. For a thin disk,
the material supply in this way accretes onto the BH as a power-law in
time with index $-1.2$ to $-1.25$, according to the viscosity
prescription \cite{fkr}. A solid line with slope $-1.25$ is shown in
Fig.~\ref{fig:curve}. Such a possibility is favored by the detection
of flares in short GRBs (without an associated supernova nor a
progenitor star). However, not enough flares in short GRBs have been
observed to meaningfully check whether or not they have the same
properties of those observed in long GRBs. An alternative scenario is
provided by an inner engine made by a fast spinning magnetar
\cite{buccia08}. In that case, though, the spin down luminosity should
be steeper than allowed by the data.

\section{From steady state to flares}

\begin{figure}
\includegraphics[width=0.5\textwidth]{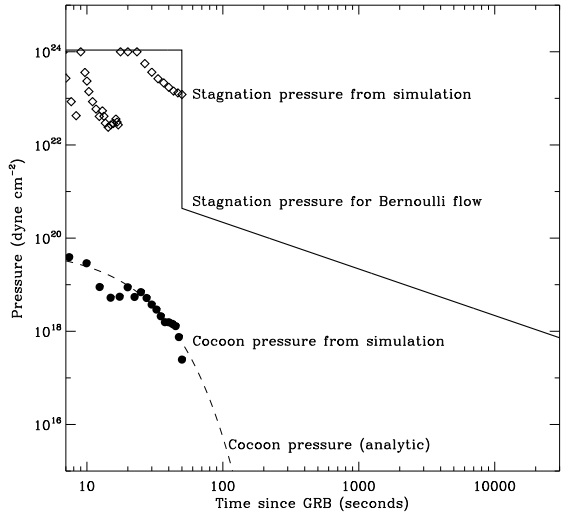}
\caption{{Comparison between the cocoon pressure and the jet 
stagnation pressure for a typical GRB jet penetrating a 15 solar
masses star. Lines show analytical estimates from \cite{lazzbege05},
while symbols show simulation results from \cite{brian07}.}
\label{fig:pstag}}
\end{figure}

Figure~\ref{fig:curve} shows a continuous light curve. However, flares
are episodic events, and we have to explain not only where the energy
is taken from, but also the reason why its release is
discontinuous. One possibility is that a continuous outflow is
converted into a discontinuous one by the interaction with the stellar
progenitor material. This would be possible if the pressure of the
cocoon surrounding the jet inside the star overcomes, at any time, the
stagnation pressure of the jet:
\begin{equation}
p_{\rm{stag}} = \frac{L_j\,\Gamma^2}{4c\Sigma}
\label{eq:pstag}
\end{equation}

Figure~\ref{fig:pstag} shows the comparison between the stagnation
pressure of a typical GRB jet and the cocoon pressure. Lines (solid
and dashed) show the result of analytic estimates based on
\cite{lazzbege05}. The result is that the cocoon pressure is never
strong enough to actually stagnate the jet. This analytic estimate is
supported by simulation results from \cite{brian07}, shown in the
figure with symbols. Even though the jet stagnation pressure can
decrease due to shocking and non-adiabatic evolution, it hardly gets
close to the cocoon pressure. The consequence of this is that the
flaring nature of the late energy release from the inner engine must
be related either to the accretion process itself \cite{perna06} or
the the jet launching process, before the outflow becomes
relativistic.

\section{Conclusions}
We have analyzed X-ray flares in long duration GRB afterglows.We
reached several conclusions. First, they are due to activity of the
engine at the same time they are observed. This allowed us to produce
a light curve of the energy emitted by the engine at late times. We
find that it falls as a power-law with index $\sim-1.5$. the data are
consistent with two processes to power the flares: fallback accretion
and disk draining. Finally, we excluded that the flaring behavior can
be due to the propagation of the relativistic outflow through the
star. It has to be due either to the accretion process or to the jet
launching mechanism.

\begin{theacknowledgments}
This work was supported by NSF grants AST-0307502 and AST-0507571,
NASA ATP grant NNG06GI06G, and Swift GI program NNX06AB69G.
\end{theacknowledgments}

\bibliographystyle{aipprocl} 
\bibliography{lazzati}

\begin{thebibliography}{10}
\providecommand{\enquote}[1]{``#1''}
\expandafter\ifx\csname url\endcsname\relax
  \def\url#1{\texttt{#1}}\fi
\expandafter\ifx\csname urlprefix\endcsname\relax\def\urlprefix{URL }\fi

\bibitem{chinca07}
G.~{Chincarini}, A.~{Moretti}, P.~{Romano}, A.~D. {Falcone}, D.~{Morris},
  J.~{Racusin}, S.~{Campana}, S.~{Covino}, C.~{Guidorzi}, G.~{Tagliaferri},
  D.~N. {Burrows}, C.~{Pagani}, M.~{Stroh}, D.~{Grupe}, M.~{Capalbi},
  G.~{Cusumano}, N.~{Gehrels}, P.~{Giommi}, V.~{La Parola}, V.~{Mangano},
  T.~{Mineo}, J.~A. {Nousek}, P.~T. {O'Brien}, K.~L. {Page}, M.~{Perri},
  E.~{Troja}, R.~{Willingale}, and B.~{Zhang}, \emph{ApJ} \textbf{671},
  1903--1920 (2007).

\bibitem{falcone07}
A.~D. {Falcone}, D.~{Morris}, J.~{Racusin}, G.~{Chincarini}, A.~{Moretti},
  P.~{Romano}, D.~N. {Burrows}, C.~{Pagani}, M.~{Stroh}, D.~{Grupe},
  S.~{Campana}, S.~{Covino}, G.~{Tagliaferri}, R.~{Willingale}, and
  N.~{Gehrels}, \emph{ApJ} \textbf{671}, 1921--1938 (2007).

\bibitem{lazzperna07}
D.~{Lazzati}, and R.~{Perna}, \emph{MNRAS} \textbf{375}, L46--L50 (2007).

\bibitem{lazz08}
D.~{Lazzati}, R.~{Perna}, and B.~M. C., \emph{MNRAS submitted}  (2008).

\bibitem{cheva}
R.~A. {Chevalier}, \emph{ApJ} \textbf{346}, 847--859 (1989).

\bibitem{fkr}
J.~{Frank}, A.~{King}, and D.~J. {Raine}, \emph{{Accretion Power in
  Astrophysics: Third Edition}}, Accretion Power in Astrophysics, by Juhan
  Frank and Andrew King and Derek Raine, pp.~398.~ISBN 0521620538.~Cambridge,
  UK: Cambridge University Press, February 2002., 2002.

\bibitem{buccia08}
N.~{Bucciantini}, E.~{Quataert}, J.~{Arons}, B.~D. {Metzger}, and T.~A.
  {Thompson}, \emph{MNRAS} \textbf{383}, L25--L29 (2008).

\bibitem{lazzbege05}
D.~{Lazzati}, and M.~C. {Begelman}, \emph{ApJ} \textbf{629}, 903--907 (2005).

\bibitem{brian07}
B.~J. {Morsony}, D.~{Lazzati}, and M.~C. {Begelman}, \emph{ApJ} \textbf{665},
  569--598 (2007).

\bibitem{perna06}
R.~{Perna}, P.~J. {Armitage}, and B.~{Zhang}, \emph{ApJL} \textbf{636},
  L29--L32 (2006).

\end{thebibliography}

\end{document}